\documentclass[11pt,showpacs,preprintnumbers,amsmath,amssymb]{article}

\usepackage{graphicx}
\usepackage{dcolumn}
\usepackage{bm}
\usepackage{amssymb}
\usepackage{amsmath}

\def\bear{\begin{eqnarray}}
\def\ear{\end{eqnarray}}

\def\no{\noindent}
\def\e{{\rm e}}

\begin{document}

\hfill \medskip NSF-KITP-14-116

\begin{center}

\centerline{\Large\bf On the Vlasov equation for Schwinger pair production}

\centerline{\Large\bf in a time-dependent electric field}

\bigskip
\bigskip

{Adolfo Huet$^{a,b}$, Sang Pyo Kim$^{c,d}$, Christian Schubert$^{b,e}$}

\begin{itemize}
\item [$^a$]
{\it
Departamento de Nanotecnolog{{\'\i}}a,
Centro de F{{\'\i}}sica Aplicada y Tecnolog{{\'\i}}a Avanzada, (CFATA)
UNAM, Campus Juriquilla, Boulevard Juriquilla No. 3001, C.P. 76230, A.P. 1-1010, Quer\'etaro, M\'exico.
E-mail: ahuet@fata.unam.mx.
 }
 \item [$^b$]
{\it 
Instituto de F\'{\i}sica y Matem\'aticas, 
Universidad Michoacana de San Nicol\'as de Hidalgo,
Edificio C-3, Apdo. Postal 2-82,
C.P. 58040, Morelia, Michoac\'an, M\'exico.
E-mail: schubert@ifm.umich.mx.
}

  \item[$^c$]
  {\it 
Department of Physics, Kunsan National University, Kunsan 573-701, Korea.
E-mail: sangkim@kunsan.ac.kr.
}
\item [$^d$] 
 {\it 
Center for Relativistic Laser Science, Institute for Basic Science, Gwangju 500-712, Korea.
}
\item [$^e$]
{\it
Kavli Institute for Theoretical Physics,
University of California, 
Santa Barbara, CA  93106, USA.
}
\end{itemize}
\end{center}

\bigskip

\date{\today}

\begin{abstract}
Schwinger pair creation in a purely time-dependent electric field can be described through a quantum Vlasov
equation describing the time evolution of the single-particle momentum distribution function. This equation
exists in two versions, both of which can be derived by a Bogoliubov transformation, but  whose
equivalence is not obvious. For the spinless case, we show here that the difference between these two evolution 
equations corresponds to the one between the ``in-out'' and ``in-in'' formalisms. We give a simple relation between the
asymptotic distribution functions generated by the two Vlasov equations. As examples we discuss the Sauter 
and single-soliton field cases.

\end{abstract}


\vfill\eject

\section{Introduction}\label{introduction}

The QED effect of spontaneous pair creation of electron-positron pairs by a strong electric field was conceived by Sauter 
as early as 1931 \cite{sauter} and computed by Schwinger in 1951 for the constant field case \cite{schwinger51}.
Considerable work has gone into developing techniques for the calculation of this effect for
more realistic electric field configurations \cite{breitz,popov,bggs,marpop,kescm,gavgit,klmoei,kimpage,dipiazza,giekli,63,64,dumdun}, 
in particular such as can be realized by laser fields (see, e.g., \cite{ringwald,dunne_laser,tajima}).
In this effort, a special role is played by the case of a purely time-dependent field. This is because, although even here
analytic results cannot be expected in the generic case, a formalism exists that allows one to compute the pair creation
rate numerically in a straightforward manner. 
What makes the purely time-dependent case very special is that for such a field the spatial momentum $\bf k$ is a good quantum number. 
Thus one can fix it,   and use a Bogoliubov transformation between the vacua at initial time and time $t$  
to derive an evolution equation for the density of pairs ${\cal N}_{\bf k}(t)$ with fixed momenta. 
This is the quantum Vlasov equation (`QVE'). In the scalar QED case, which we will consider in this paper,
the QVE is usually written in the form \cite{kescm,klmoei,healgi,srsbtp,schton,sbrpst,bmprssv,ahrsv,dumlu}

\begin{eqnarray}
\dot{\cal N}_{\bf k} (t)
&=& \frac{\dot\omega_{\bf k} (t)}{2\omega_{\bf k}(t)} \int_{-\infty}^{t} dt' \, \frac{\dot\omega_{\bf k} (t')}{\omega_{\bf k}(t')}
(1+2 {\cal N}_{\bf k} (t') ) \nonumber\\
&&\times  \cos \Bigl\lbrack 2\int_{t'}^{t} dt'' \omega_{\bf k} (t'') \Bigr\rbrack  
\label{inout}
\end{eqnarray}
where $\omega_{\bf k}^2 (t)$ is the total energy squared,

\begin{eqnarray}
\omega_{\bf k}^2 (t) &=& (k_{\parallel} - qA_{\parallel} (t))^2 + {\bf k}_{\perp}^2 + m^2 \, .
\label{defomega}
\end{eqnarray}
Here the temporal gauge $A_0=0$,  $\dot {\bf A}(t) = - {\bf E}(t)$ has been used. The field points into a fixed direction,
and $k_{\parallel}$ denotes the canonical three-momentum component along the field. 
${\cal N}_{\bf k}(t)$ is usually taken to be zero initially, that is at $ t = -\infty$, and for $t\to\infty$ turns into the density of created pairs
with fixed momentum $\bf k$. This interpretation has been shown to be consistent with various other approaches to the time-dependent
Schwinger problem, such as the one-loop effective action \cite{healgi},  the quantum mechanical scattering approach \cite{dumlu}
and the Wigner formalism \cite{devava,elzhei,bigora,healgi2}.  From a practical point of view,  although the structure of (\ref{inout}) is not promising for attempts at
exact analytical solution, it is amenable to numerical evaluation, as well as to other approximation schemes \cite{healgi,sbrpst,bmprssv,ahrsv,dumlu}.
Sometimes it is preferable to use the following equivalent system of first-order differential equations \cite{bmprssv}

\begin{eqnarray}
\frac{d}{dt} \begin{pmatrix}
1+2{\cal N}_{\bf k} \\
{\cal M}^{(-)}_{\bf k}\\
{\cal M}^{(+)}_{\bf k}
\end{pmatrix} = \begin{pmatrix}
  0 &   \frac{\dot\omega_{\bf k} }{\omega_{\bf k}}  & 0 \\
\frac{\dot\omega_{\bf k} }{\omega_{\bf k}}     & 0 &  -2\omega_{\bf k}\\
     0& 2\omega_{\bf k} &0 \end{pmatrix} \begin{pmatrix}
1+2{\cal N}_{\bf k} \\
{\cal M}^{(-)}_{\bf k}\\
{\cal M}^{(+)}_{\bf k}
\end{pmatrix}.\nonumber\\ \label{DGLinout}
\end{eqnarray}
where ${\cal M}^{(\pm )}_{\bf k}$ are two auxiliary functions.

As has been  emphasized in \cite{healgi,fgks}, no direct physical meaning should be ascribed to ${\cal N}_{\bf k}(t)$ at intermediate times, due to its 
dependence on the choice of an instantaneous adiabatic basis. This implies a large ambiguity which one can try to use for the
construction of simpler, but physically equivalent evolution equations. 
In \cite{84} two of the authors obtained, also using a Bogoliubov transformation but in a form inspired by
Lewis-Riesenfeld theory \cite{lewrie,mmt,cdms,ffvv}, the following alternative evolution equation (see also the recent \cite{dabdun} for a systematic approach
to the construction of a mathematically optimized adiabatic basis)

\begin{eqnarray}
{\dot{\tilde{\cal N}}}_{\bf k} (t)
&=& \frac{1}{2}\Omega_{\bf k}^{(-)} (t) \int_{t_0}^{t} dt' \Bigl[ \Omega_{\bf k}^{(-)}(t')
(1+2\tilde  {\cal N}_{\bf k} (t') ) \nonumber\\
&&\times  \cos\Bigl( \int_{t'}^{t} dt'' \Omega_{\bf k}^{(+)} (t'') \Bigr) \Bigr]. \label{inin}
\end{eqnarray}
Here $\tilde{\cal N}_{\bf k}(t) = 0$ at $t=-\infty$ as before, and

\begin{eqnarray}
\Omega^{(\pm)}_{\bf k} (t) := \frac{\omega_{\bf k}^2 (t) \pm \omega_{{\bf k}i}^2 }{\omega_{{\bf k}i} }
\label{defOmega}
\end{eqnarray}
where

\bear
\omega_{{\bf k}i} := \lim_{t\to -\infty}\omega_{\bf k}(t) \, .
\label{defomegaminusinf}
\ear
Thus we have to assume in the following that this limit is finite, and similarly also

\bear
\omega_{{\bf k}f} := \lim_{t\to +  \infty}\omega_{\bf k}(t) \, .
\label{defomegaplusinf}
\ear
However, this is a minor restriction since, for electric fields that vanish sufficiently fast at $t = \pm \infty$, it can always be
achieved in a suitable gauge. 

Note that, in contrast to the coefficient functions appearing in the ``standard'' Vlasov equation (\ref{inout}), 
the coefficient functions $\Omega^{(\pm)}_{\bf k} (t)$ do not involve square roots of time-dependent quantities.

For the ``alternative'' Vlasov equation, two equivalent differential equations were obtained in \cite{84}.
The first one is a system of first-order differential equations analogous to (\ref{DGLinout}),

\begin{eqnarray}
\frac{d}{dt} \begin{pmatrix}
1+2\tilde{\cal N}_k \\
\tilde{\cal M}^{(-)}_k\\
\tilde{\cal M}^{(+)}_k
\end{pmatrix} = \begin{pmatrix}
  0 &  \Omega^{(-)}_k & 0 \\
     \Omega^{(-)}_k & 0 &  \Omega^{(+)}_k\\
     0& - \Omega^{(+)}_k &0 \end{pmatrix} \begin{pmatrix}
1+2\tilde{\cal N}_k \\
\tilde{\cal M}^{(-)}_k\\
\tilde{\cal M}^{(+)}_k
\end{pmatrix}.\nonumber\\ \label{DGLinin}
\end{eqnarray}
The second one is the following linear third order linear differential equation

\begin{eqnarray}
\dddot F + 4\omega^2\dot F+2{(\omega^2)}\dot{\phantom{.}} F
 =\frac{(\omega^2)\dot{\phantom{.}}}{\omega_i^2}
\label{DGL}
\end{eqnarray}
where the relation between $F(t)$ and $\tilde{\cal N} (t)$ is given by
(here and in the following we often drop the subscript $\bf k$)

\begin{eqnarray}
\tilde{\cal N} (t) &=& \frac{\omega_i}{2} \int_{-\infty}^t dt' \dot F(t') \Omega^{(-)}(t') \nonumber\\
\label{FtocalN}
\end{eqnarray}
(our $\omega_i$ corresponds to $\omega_0$  in \cite{84}).

For the alternative Vlasov equation, an interesting set of explicit solutions could be found 
in terms of the well-known solitons of the Korteweg-de Vries equation \cite{84,kim,99}.

Mathematically  the two evolution equations (\ref{inout}) and (\ref{inin}) are clearly not equivalent, 
and it remained an open problem to elucidate the physical relation between them.
The purpose of the present paper is to show  that the difference between them corresponds
precisely to the one between the ``in-in'' and ``in-out'' formalisms. 
This will also allow us to find a simple relation between the asymptotic distribution functions 
${\cal N}(t=\infty)$ and $\tilde{\cal N}(t=\infty)$. 
We carry this through in chapters \ref{sec:inout} and \ref{sec:inin}. 
As examples, in chapter \ref{sec:examples} we
discuss the Sauter  and single-soliton field cases. 

\section{The in-out Vlasov equation}
\label{sec:inout}

In this chapter we will verify explicitly that the ``standard'' Vlasov equation corresponds to the ``in-out'' formalism.
The argument essentially follows \cite{fgks}.  The basic equation underlying both evolutions equations is
the classical mode equation

\begin{eqnarray}
\ddot{\phi} (t) + \omega^2 (t) \phi (t) = 0\, . \label{mod eq}
\end{eqnarray}
Assuming that $\phi$ fulfills the mode equation (\ref{mod eq}), then

\bear
{\cal N}(t) := \frac{|\dot \phi (t)|^2 + \omega^2(t) |\phi(t) |^2}{2\omega(t)} - \frac{1}{2}
\label{relNphi}
\ear
solves the Vlasov equation (\ref{inout}). This can be verified most directly by plugging
${\cal N}(t)$ into the equivalent differential equation (\ref{DGLinout}), where the two auxiliary
functions ${\cal M}^{(\pm )}$ are given by

\begin{eqnarray}
{\cal M}^{(+)}(t) &=& - \frac{d}{dt} |\phi(t)|^2 \, ,  \nonumber\\
{\cal M}^{(-)}(t) &=& -  \frac{|\dot \phi (t)|^2 - \omega^2(t) |\phi(t) |^2}{\omega(t)}  \, . \nonumber\\
\label{Mpmexplicit}
\ear
Now, under our assumption 
that  $\omega_i:= \lim_{t\to -\infty}\omega(t)$ converges, we can decompose the mode solution for
$t\to -\infty$ as 

\bear
\phi(t)\, \stackrel{t\to -\infty}{\longrightarrow}\, A_0 \,\frac{\e^{i\omega_it}}{\sqrt{2\omega_i}} + B_0 \,\frac{\e^{-i\omega_i t}}{\sqrt{2\omega_i}}
\label{phiminusinf}
\ear
leading to 

\bear
{\cal N}(t) &\, \stackrel{t\to -\infty}{\longrightarrow}\, &  \frac{1}{2}(|A_0|^2+|B_0|^2) - \frac{1}{2} \, . \nonumber\\
\nonumber\\
\label{Nminusinf}
\ear
If we further assume that initially there are no particles present, standard quantum field
theory arguments imply that $A_0 = 0$, and ${\rm lim}_{t\to -\infty}{\cal N}(t)=0$
together with (\ref{Nminusinf}) then shows that $|B_0|^2 = 1$.
Thus the relevant solution of the mode equation will obey (up to an irrelevant phase factor)

\bear
\phi(t)\, \stackrel{t\to -\infty}{\longrightarrow}\, \frac{1}{\sqrt{2\omega_i}} \,\e^{-i\omega_i t} \, .
\label{condin}
\ear
At future infinity, in the in-out formalism we have to expand the mode solution in terms of plane waves with the final frequency,

\bear
\phi(t)\, \stackrel{t\to +\infty}{\longrightarrow}\, A_{\infty} \,\frac{\e^{i\omega_ft}}{\sqrt{2\omega_f}} + B_{\infty} \,\frac{\e^{-i\omega_f t}}{\sqrt{2\omega_f}} \, .
\label{phiplusinf}
\ear
Similarly to (\ref{Nminusinf}) this yields 

\bear
{\cal N}_{\infty} &:=& 
\lim_{t\to\infty}  {\cal N}(t) = \frac{1}{2}(|A_{\infty}|^2+|B_{\infty}|^2) - \frac{1}{2}  \, .
 \nonumber\\
\nonumber\\
\label{Nplusinf}
\ear
A further relation is obtained from the Wronskian constraint (see, e.g., \cite{84})

\begin{eqnarray}
{\rm Wr} [\phi, \phi^*] \equiv \phi (t) \dot{\phi}^* (t) - \phi^* (t) \dot{\phi} (t) = i \label{wronski}
\end{eqnarray}
which gives

\bear
|B_{\infty}|^2-|A_{\infty}|^2 = 1 .
\label{BtoA}
\ear
Using this equation in (\ref{Nplusinf}) to eliminate $B_{\infty}$ we obtain the simple expression

\bear
{\cal N}_{\infty} &=& |A_{\infty}|^2 
\nonumber\\
\label{Ninfty}
\ear 
for the final density of created pairs (at fixed momentum ${\bf k}$).

\section{The in-in Vlasov equation}
\label{sec:inin}

We will now proceed analogously for the alternative Vlasov equation (\ref{inin}). 
Here the analogue of (\ref{relNphi}), that is, an explicit solution of  (\ref{inin}) in terms of the solution of the mode equation (\ref{mod eq}), 
turns out to be

\bear
\tilde {\cal N}(t) = \frac{|\dot \phi (t)|^2 + \omega_i^2 |\phi(t) |^2}{2\omega_i} - \frac{1}{2} 
\label{reltildeNphi}
\ear
as can again be verified simply by plugging into the  differential equation
system (\ref{DGLinin}). The auxiliary
functions $\tilde{\cal M}^{(\pm )}$ are now given by

\begin{eqnarray}
\tilde{\cal M}^{(+)} &=&  -  \frac{|\dot \phi (t)|^2 - \omega_i^2 |\phi(t) |^2}{\omega_i} \, ,  \nonumber\\
\tilde{\cal M}^{(-)} &=&  - \frac{d}{dt} |\phi(t)|^2   \nonumber\\
\label{tildeMpmexplicit}
\ear
(note the interchange of $(+)$ and $(-)$ as compared to (\ref{Mpmexplicit})).
The corresponding solution of the third-order differential equation (\ref{DGL}) is

\bear
F(t) = -\frac{1}{\omega_i} \Bigl(|\phi(t) |^2 - \frac{1}{2\omega_i} \Bigr) \, .
\label{relFphi}
\ear
For the sake of completeness, in the appendix we show how to derive this solution, as well as the equation
itself, directly from Lewis-Riesenfeld theory.

Since (\ref{reltildeNphi}) differs from (\ref{relNphi}) only by the replacement of $\omega(t)$ by $\omega_i$, 
the asymptotic expansion of $\tilde {\cal N}(t)$ for $t\to -\infty$ agrees with the one of ${\cal N}(t)$, eq.
(\ref{Nminusinf}). However, at large times things are very different. 
To get formulas analogous to (\ref{phiplusinf}), (\ref{Nplusinf}) we now
have to project on plane waves with the {\it initial} frequencies. Namely, defining $ \tilde A_{\infty}, \tilde B_{\infty}$ by

\bear
\phi(t)\, \stackrel{t\to +\infty}{\longrightarrow}\, \tilde A_{\infty} \,\frac{\e^{i\omega_it}}{\sqrt{2\omega_i}} + \tilde B_{\infty} \,\frac{\e^{-i\omega_i t}}{\sqrt{2\omega_i}}
\label{phiplusinfalt}
\ear
eq. (\ref{reltildeNphi}) gives

\bear
\tilde{\cal N}(t) &\, \stackrel{t\to +\infty}{\longrightarrow}\, &  \frac{1}{2}(|\tilde A_{\infty}|^2+|\tilde B_{\infty}|^2) - \frac{1}{2} \, . \nonumber\\
\nonumber\\
\label{Ntildeplusinfalt}
\ear
Using also the Wronskian constraint to eliminate $\tilde B_{\infty}$, we find

\bear
\tilde{\cal N}_{\infty} &:=& \lim_{t\to \infty} \tilde{\cal N}(t) = |\tilde A_{\infty}|^2.
\nonumber\\
\label{tildeNinfty}
\ear 
Thus $\tilde {\cal N}$ asymptotically measures the presence of negative frequency components with respect to the {\it initial} energies.
This provides an explicit way of seeing that the alternative Vlasov equation (\ref{inin}) pertains  to the in-in formalism. 

To be able to transform between the two equations we need to know also how $\tilde {\cal N}(t)$ is written asymptotically in terms
of the coefficients $A_{\infty},B_{\infty}$ of the expansion of $\phi$ in terms of {\it final} frequencies, eq. (\ref{phiplusinf}). Here we find

\bear
\tilde{\cal N}(t) &\, \stackrel{t\to +\infty}{\longrightarrow}\, & \frac{\omega_i^2+\omega_f^2}{4\omega_i\omega_f} (|A_{\infty}|^2+|B_{\infty}|^2) -\frac{1}{2} 
\nonumber\\&&
+ \frac{\omega_i^2-\omega_f^2}{4\omega_i\omega_f} (A_{\infty}B_{\infty}^{\ast}\,\e^{2i\omega_ft} + A_{\infty}^{\ast}B_{\infty}\,\e^{-2i\omega_ft}) \, .
\nonumber\\
\label{Ntildeplusinf}
\ear
Comparing with (\ref{Nplusinf}) we see that ${\cal N}(t)$ and $\tilde {\cal N}(t)$ will agree for $t\to \infty$ if and only if $\omega_f=\omega_i$. 
Otherwise $\tilde{\cal N}(t)$ will not even converge, due to the oscillating terms in the second line of (\ref{Ntildeplusinf}).
To eliminate these terms, we can use asymptotic averaging. Defining the asymptotic average of a function $f(t)$ for large $t$ as usual by

\bear
\langle f \rangle_{\infty} := \lim_{t\to\infty} \frac{1}{t}\int_t^{2t}dt' f(t')
\ear
we get the transformation formula

\bear
2{\cal N}_{\infty}+ 1 = 2\frac{\omega_i\omega_f}{\omega_i^2+\omega_f^2} \langle 2\tilde {\cal N} + 1 \rangle_{\infty} \, .
\label{trafo}
\ear

\section{Examples}
\label{sec:examples}

We will illustrate the differences between ${\cal N}(t)$ and $\tilde{\cal N}(t)$ with two examples for which closed-form
solutions of the mode equation (\ref{mod eq}) exist, namely the ``time-like Sauter field'' and the ``single-soliton field''.
The time-like Sauter field is defined by

\bear
E(t) = E_0 \,{\rm sech}^2(t/\tau)
\ear
which we can realize by 

\bear
A^{\mu} = (0,0,0,E_0 \tau (1+\tanh (t/\tau) )) \, .
\ear
The solution to the mode equation obeying (\ref{condin}) is \cite{kileyo}

\bear
\phi(t) &=& \frac{1}{\sqrt{2\omega_0 \e^{\pi\omega_o\tau}}}\Bigl(-\e^{\frac{2t}{\tau}}\Bigr)^{-\frac{i}{2}\omega_0\tau}\Bigl(1+\e^{\frac{2t}{\tau}}\Bigr)^{\frac{1}{2}+i\sqrt{-\frac{1}{4}+E_0^2\tau^4}}
\nonumber\\
&&\times
\phantom{}_2F_1\Bigl(\frac{1}{2} - \frac{i}{2}\Bigl(\omega_0\tau -2\sqrt{-\frac{1}{4}+E_0^2\tau^4}-\tau\sqrt{\omega_0^2 - 4E_0\tau +4E_0^2\tau^2}\Bigr),
\nonumber\\&&
\frac{1}{2} - \frac{i}{2}\Bigl(\omega_0\tau -2\sqrt{-\frac{1}{4}+E_0^2\tau^4}+\tau\sqrt{\omega_0^2 - 4E_0\tau +4E_0^2\tau^2}\Bigr),
1-i\omega_0\tau,-\e^{\frac{2t}{\tau}} \Bigr)
\, .
\nonumber\\
\label{phisauter}
\ear
where $\omega_0 \equiv \sqrt{{\bf k}^2+m^2}$, and the gauge has been chosen such that $\omega_i=\omega_0$. 
For our example, we use the parameters $E_0=q=k_{\parallel}=1,\omega_0=1.1,\tau = 2$. Plugging $\phi(t)$ into (\ref{relNphi}) and (\ref{reltildeNphi})
results in expressions for ${\cal N}(t), \tilde{\cal N}(t)$ which are too lengthy to be given here. A MATHEMATICA plot of both function for the our parameter values is 
shown in fig. \ref{fig_sautercompare}. Here $\omega_f = 3.04 \ne \omega_i= \omega_0=1.1$, therefore as stated
above $\tilde{\cal N}(t)$ keeps oscillating for $t\to\infty$. It is easily checked that the transformation formula (\ref{trafo}) between ${\cal N}_{\infty}$ 
and $\langle \tilde{\cal N} \rangle_{\infty}$ is fulfilled.

\begin{figure}[htp]
\vspace{42pt}
\center{\includegraphics[width=7.0cm]{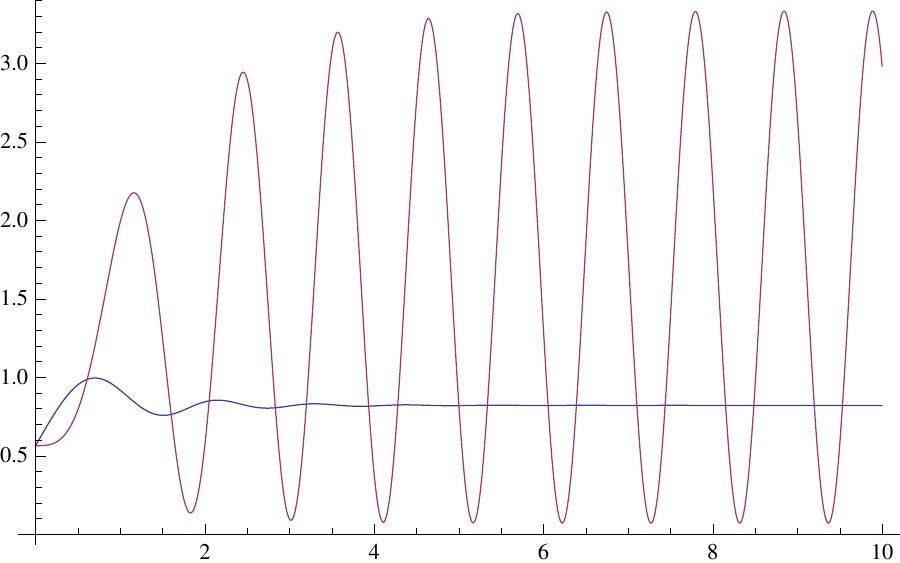}}
\caption{A comparison of ${\cal N}(t)$ (damped oscillating blue curve) against $\tilde{\cal N}(t)$ (oscillating red curve) for the timelike Sauter field.}
\label{fig_sautercompare}
\end{figure}

Our second example is a ``solitonic'' electric field. In \cite{84,kim} it had been shown that an infinite set of special solutions to the
alternative Vlasov equation can be explicitly constructed using the well-known soliton solutions of the Korteweg-de-Vries equation.
The simplest one of these field configurations is given by 

\begin{eqnarray}
qA(t) = k_{\parallel} - \sqrt{k_{\parallel}^2+ \frac{2\omega_0^2}{\cosh^2(\omega_0 t)}}
\label{defAsoliton}
\end{eqnarray}
where again $\omega_i = \omega_0$. The solitonic character of the associated solutions
shows itself in the property that ${\rm lim}_{t\to\infty} {\cal N}(t) = 0$ for the $\bf k$ used in the
definition of the field (\ref{defAsoliton}). This does not mean that such an electric field does not
pair-create at all,  but it can be tuned not to pair-create at a given momentum. 
The solution to the mode equation obeying (\ref{condin}) is \cite{84}

\bear
\phi(t) = \frac{1}{\sqrt{2\omega_0}} \frac{\e^{-i\omega_0 t}(1-i\e^{2\omega_0 t})}{1+\e^{2\omega_0 t}} \, .
\label{phisoliton}
\ear
Plugging this into (\ref{relNphi}) reps. (\ref{reltildeNphi}) gives

\begin{eqnarray}
{\cal N}(t) &=& \frac{4+{\rm sech}^4(\omega_0 t)(1+2\cosh(2\omega_0 t))}{8\sqrt{1+2{\rm sech}^2(\omega_0 t)}}- \frac{1}{2} \, ,
\nonumber\\
\tilde{\cal N}(t) &=& \frac{1}{8\cosh^4(\omega_0 t)} \, .
\nonumber\\
\label{NNbarsoliton}
\end{eqnarray}
In fig. \ref{fig_solitoncompare} we show a plot of both functions ${\cal N}(t), \tilde{\cal N}(t)$. 
Note the simplicity of $\tilde{\cal N}(t)$, the
symmetry with respect to $t=0$, the absence of oscillations and the vanishing of both functions for
$t\to\infty$. 

\begin{figure}[htp]
\vspace{-5pt}
\hspace{11.5pt}
\center\includegraphics[width=6.0cm]{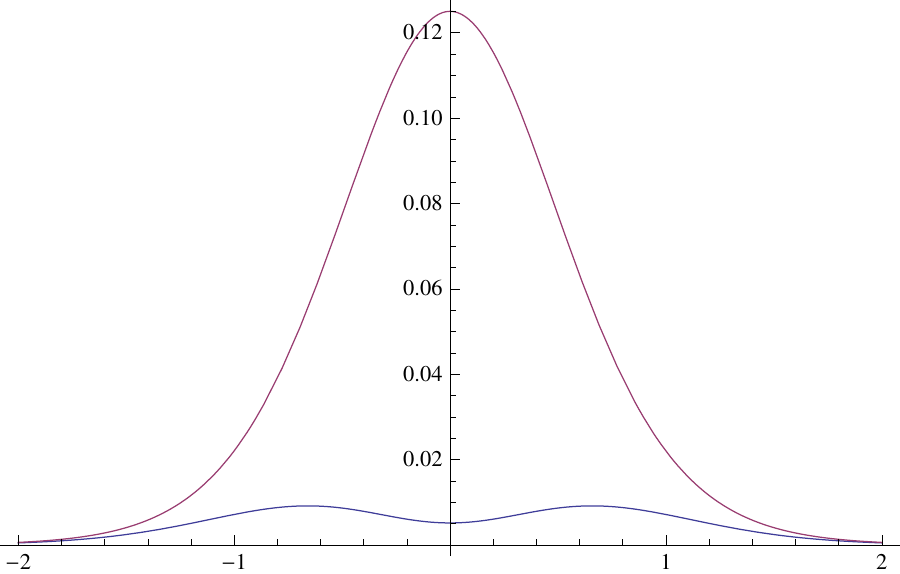}
\caption{A comparison of ${\cal N}(t)$ (double peaked blue curve) against $\tilde{\cal N}(t)$ (single peaked red curve) for the simplest solitonic field.}
\label{fig_solitoncompare}
\end{figure}

\section{Conclusions} 
\label{conclusions}

We have clarified here the relation of the Vlasov equation for the creation of scalar pairs by time-dependent fields,
recently proposed in \cite{84}, to the older and widely used one of \cite{klmoei}. 
The difference between both equations boils down to a projection on final states with initial vs. final frequencies, that is, 
just to the distinction between the ``in-in'' vs. the ``in-out'' formalisms. 
If the initial and final frequencies are equal, $\omega_f =\omega_i$, the asymptotic pair production rates coincide. This is the case, in particular,
for the family of ``solitonic'' fields found in \cite{84,kim}. Thus the property of the associated solutions
of the alternative Vlasov equation to have a vanishing asymptotic pair creation rate is shared by the corresponding
solutions of the standard Vlasov equation.

 If $\omega_f \ne \omega_i$ the latter equation is, as the in-out one, the appropriate one
in the context of laboratory experiments (the in-in formalism figures prominently in cosmology, see, e.g., \cite{himamo,moss}). 
In any case, with the transformation formula (\ref{trafo}) in hand both equations can be used
equivalently according to mathematical convenience.  

\bigskip
\bigskip

\noindent
{\bf Acknowledgements:}
The authors would like to thank 
R. Alkofer, G.V. Dunne, A.~M. Fedotov, H. Gies,  F. Hebenstreit and S.-J. Rey for discussions and correspondence. 
C.~S.~ thanks  the KITP for hospitality during the program `Frontiers of intense laser physics', as well as the
Physics Department of Kunsan National University, Korea, and the
Center for Theoretical Physics, Seoul National University.  
The work of A. Huet and C. Schubert was supported by CONACYT, Mexico,
the work of S.~P. K. by the Research Center Program of IBS (Institute for Basic Science, Korea)
under IBS-R012-D1, and the work of C. S. by the NSF under Grant No. NSF PHY11-25915.

\bigskip
\bigskip

\noindent
{\bf Appendix}

\medskip
\no
For completeness, in this appendix we provide a direct derivation of the third-order equation (\ref{DGL}), and its solution,
using the Lewis-Riesenfeld approach to QED of \cite{84}.
To avoid undue repetition, in this appendix we refer the reader to \cite{84} for the general set-up and notation. 
The Hermitian conjugates of the fields in (6) of \cite{84} are

\begin{eqnarray}
\hat{\phi}^{\dagger}_{\bf k} = \hat{a}^{\dagger}_{\bf k} (t) \phi^*_{\bf k} (t) + \hat{b}_{- \bf k} (t) {\phi}_{\bf k} (t), \nonumber\\
\hat{\pi}^{\dagger}_{\bf k} = \hat{a}_{\bf k} (t) \dot\phi_{\bf k} (t) + \hat{b}^{\dagger}_{- \bf k} (t) \dot{\phi}^*_{\bf k} (t), \nonumber\\
\end{eqnarray}
where the quantum invariants $\hat{a}_{\bf k} (t), \hat{a}^{\dagger}_{\bf k} (t) $ and $ \hat{b}_{- \bf k} (t), \hat{b}^{\dagger}_{- \bf k} (t)$ are the time-dependent annihilation and creation operators for particle and antiparticle, respectively, and the auxiliary field $\phi_{\bf k}$ obeys  (\ref{mod eq}). The vacuum expectation values of the equal-time correlation functions are
\begin{eqnarray}
\langle \hat{\phi}^{\dagger}_{\bf k} \hat{\phi}_{\bf k} \rangle &=& |\phi_{\bf k} (t)|^2, \nonumber\\
\langle \hat{\pi}^{\dagger}_{\bf k} \hat{\phi}^{\dagger}_{\bf k} + \hat{\phi}_{\bf k} \hat{\pi}_{\bf k}\rangle &=& \frac{d}{dt}|\phi_{\bf k} (t)|^2, \nonumber\\
\langle \hat{\pi}^{\dagger}_{\bf k} \hat{\pi}_{\bf k} \rangle &=& |\dot{\phi}_{\bf k} (t)|^2. \nonumber\\
\label{3 exp} 
\end{eqnarray}
The mode solution may be written as (now dropping the subscript ${\bf k}$)
\begin{eqnarray}
\phi (t) = \rho (t) e^{- i \theta (t)},
\label{phitorho}
\end{eqnarray}
which from the quantization rule (\ref{wronski}) satisfies the nonlinear equation
\begin{eqnarray}
\ddot{\rho} (t) + \omega^2 (t) \rho (t) - \frac{1}{4 \rho^3 (t)} = 0. \label{nonl mod}
\end{eqnarray}
The expectation value of the Hamiltonian (1) of  \cite{84} equated to the integral of the frequency change for the potential energy
\begin{eqnarray}
\langle \hat{H} (t) \rangle = \dot{\rho}^2 + \omega^2 \rho^2 + \frac{1}{4 \rho^2} = \int^t dt' (\omega^2)\dot{\phantom{.}} \rho^2
\end{eqnarray}
leads to (\ref{nonl mod}). We may write the expectation values (\ref{3 exp}) in terms of $\rho$:
\begin{eqnarray}
\langle \hat{\phi}^{\dagger} \hat{\phi} \rangle &=& \rho^2, \nonumber\\
\langle \hat{\pi}^{\dagger} \hat{\phi}^{\dagger} + \hat{\phi} \hat{\pi} \rangle &=& 2 \rho \dot{\rho}, \nonumber\\
\langle \hat{\pi}^{\dagger} \hat{\pi} \rangle &=& \dot{\rho}^2 + \frac{1}{4 \rho^2}. \nonumber\\
\label{3 exp-2}
\end{eqnarray}
Finally, using (\ref{phitorho}) and (\ref{nonl mod}) we obtain (\ref{relFphi}) and its first three derivatives
\begin{eqnarray}
\omega_i F &=& - \rho^2 + \frac{1}{2 \omega_i}, \nonumber\\
\omega_i \dot{F} &=& - 2 \rho \dot{\rho}, \nonumber\\
\omega_i \ddot{F} &=& - 2 \Bigl(\dot{\rho}^2 + \frac{1}{4 \rho^2} \Bigr) + 2 \omega^2 \rho^2, \nonumber\\
\omega_i \dddot{F} &=&  8 \omega^2 \rho \dot{\rho} + 2 (\omega^2)\dot{\phantom{.}} \rho^2. \nonumber\\
\label{der F}
\end{eqnarray}
Now (\ref{DGL}) follows from (\ref{der F}). Note that $F, \dot{F}$ and $\ddot{F}$ are determined by the correlation functions (\ref{3 exp-2})
and that (\ref{DGL}) is a linearization of (\ref{nonl mod}).

\end{document}